\documentclass{elsart}

\usepackage{graphicx}
\usepackage{latexsym,amssymb}

\begin{document}

\begin{frontmatter}



\title{Theory of charge fluctuations and domain relocation times in
semiconductor superlattices}


\author{L. L. Bonilla\corauthref{cor1}}
\corauth[cor1]{Corresponding Author. E-mail: bonilla@ing.uc3m.es}
\address{Departamento de Matem\'aticas, Universidad Carlos III de Madrid, Av. de
la Universidad, 30, Legan\'es 28911, Spain}


\begin{abstract}
Shot noise affects differently the nonlinear electron transport in semiconductor 
superlattices depending on the strength of the coupling among the superlattice quantum 
wells. Strongly coupled superlattices can be described by a miniband Boltzmann-Langevin 
equation from which a stochastic drift-diffusion equation is derived by means of a
consistent Chapman-Enskog method. Similarly, shot noise in weakly coupled, highly 
doped semiconductor superlattices is described by a stochastic discrete drift-diffusion model. 
The current-voltage characteristics of the corresponding deterministic model consist 
of a number of stable branches corresponding to electric 
field profiles displaying two domains separated by a domain wall. 
If the initial state corresponds to a voltage on the middle of a stable branch
and is suddenly switched to a final voltage corresponding to the next branch, 
the domains relocate after a certain delay time, called relocation time. 
The possible scalings of this mean relocation time are discussed using bifurcation theory and
the classical results for escape of a Brownian particle from a potential well. 
\end{abstract}


\end{frontmatter}

\section{Introduction}
\label{1:intro}
Nonlinear charge transport in semiconductor superlattices has been the object of 
many theoretical and experimental studies in the past decade \cite{Wacpr02,Bonjpcm02}.
Superlattices are artificial spatially periodic structures first proposed by Esaki and Tsu
in order to realize a device that exhibits Bloch oscillations \cite{ETsijrd70}.
A superlattice (SL) in its simplest form contains a large number of periods, each 
comprising two layers, which are semiconductors or insulators with different 
energy gaps, but with similar lattice constants  e.~g., GaAs and AlAs. These SLs are 
synthesized by molecular-beam epitaxy or related epitaxial growth techniques in the 
vertical direction. The conduction band edge of an infinitely long ideal SL is 
modulated so that in the vertical direction it looks like a one-dimensional (1D) crystal, 
which is formed by a periodic succession of a quantum well (GaAs) and a barrier (AlAs). 
Typical experiments of vertical charge transport use an undoped or doped SL of finite
length placed in the central part of a diode (forming a $p$-$i$-$n$ or $n^+$-$n$-$n^+$ 
structure) with respective contacts at either end of the diode. Depending on the bias condition, 
the SL configuration, the doping density, the temperature or other control parameters, the 
current through the SL and the electric field distribution inside the SL display a great variety 
of nonlinear phenomena such as pattern formation, current self-oscillations, and chaotic 
behavior \cite{Wacpr02,Bonjpcm02}.

To describe and understand nonlinear charge transport in SLs, it is essential to distinguish
between {\em weakly coupled} and {\em strongly coupled} SLs. Weakly coupled SLs
(WCSLs) contain rather thick barriers separating the SL quantum wells, i.~e., the barrier 
width is much larger than the typical electron wavelength inside the barrier. Therefore,
a description of the electronic properties of WCSLs can be based on the subband structure 
of the corresponding isolated quantum well together with resonant tunneling across the 
barrier of two adjacent wells. In contrast, the quantum wells of strongly coupled SLs 
(SCSLs) are separated by thin barriers so that the electronic properties of SCSLs can be 
described in terms of extended states such as Bloch functions. The simplest mathematical 
models applied to a SL give rise to balance equations involving mesoscopic quantities such 
as the electric field, the electron density, the drift velocity, etc. The task of deriving these 
equations from first principles is far from being completed, although reasonable
deterministic balance equations have been derived for both SCSLs \cite{BEPprb03}
and for WCSLs \cite{Bonjpcm02} in particular limiting cases. 

A fundamental difference between WCSLs and SCSLs is that the former are governed by 
{\em spatially discrete} balance equations (differential-difference equations), whereas the 
latter are governed by {\em spatially continuous} equations (partial differential equations).
Both types of equations may have solutions, whose electric field profiles display regions 
of high electric field coexisting with regions of low electric field. The resulting dynamical 
behavior is very different for these two types of equations. For SCSLs, which 
are described by continuous balance equations, the field profile consists of a
charge dipole moving with the flow of electrons, which very much resembles the Gunn effect 
in bulk semiconductors \cite{Gunssc63}. Under dc voltage bias, this basic motion 
results in self-sustained oscillations of the current through the SL due to the periodic 
movement of dipole domains. In contrast, in WCSLs, which are described by discrete 
balance equations \cite{BGCprb94,Bonjpcm02}, electric-field domains (EFDs) are 
separated by a domain wall, which consists of a charge monopole. The domain wall in a 
WCSL may move with or opposite to the electron flow, or is pinned, depending on the value 
of the current \cite{CBWpre00}. This pinning of the domain wall occurs only in the 
discrete models.

In this paper, we are interested in modeling the effects of shot noise in SLs, and then 
discussing its influence on the motion of EFDs. Shot noise is a consequence of the
quantization of the charge \cite{BBupr00}. For SCSLs, we shall describe in Section
\ref{sec:SCSL} shot-noise effects by using a miniband Boltzmann-Langevin equation, from 
which we shall derive a stochastic drift-diffusion equation in the hyperbolic limit by means
of a consistent Chapman-Enskog method. The resulting stochastic drift-diffusion equation
is similar to that describing fluctuations in the Gunn effect, and therefore enhancement of 
charge fluctuations near the threshold to self-sustained oscillations of the current is to be
expected \cite{Keibook87}. 

The discrete balance equations describing electron transport in WCSLs have not been derived 
consistently from a kinetic theory, and therefore we cannot follow a perturbative treatment 
as in the case of the SCSLs. Shot-noise effects will therefore be described in Section 
\ref{sec:WCSL} by Langevin equations resulting from adding appropriate white-noise 
sources to the discrete equations \cite{BSSprb02}. An important situation to observe the 
effects of shot noise is EFD relocation due to voltage switching \cite{LGPprb98}. The 
current-voltage characteristics of a WCSL consists of a number of stable branches 
corresponding to electric field profiles displaying two domains separated by a domain wall. 
If the initial state corresponds to a voltage on the middle of a stable branch
and is suddenly switched to a final voltage corresponding to the next branch, 
the domains relocate after a certain delay time, called relocation time. The deterministic
theory of domain relocation is based on numerical simulations of a nonlinear
differential-difference model \cite{AWBpre01}. One of the main effects of shot noise is 
to render the relocation time random. If the final voltage after switching is near the limit
point that marks the end of the initial stable solution branch, the mean relocation time
can be investigated using bifurcation theory. We calculate a possible scaling of this mean 
relocation time using a stochastic amplitude equation corresponding to the normal form of
a saddle-node plus a term due to projected shot noise. For this equation, the classical results 
for escape of a Brownian particle from a potential well yield the scaling of the mean 
relocation time. The last Section contains our conclusions.

\section{Shot-noise effects in strongly-coupled superlattices}
\label{sec:SCSL}
A deterministic simple model of one-dimensional (1D) electron transport in a SCSL 
consists of the following Boltzmann-Poisson system \cite{BEPprb03}:
\begin{eqnarray} 
&&{\partial f\over \partial t} + v(k) {\partial f\over \partial x} +  {eF\over 
\hbar} {\partial f\over \partial k} = - \nu_{e}\,  \left(f - f^{FD}\right) 
- \nu_{i}\, {f(x,k,t) - f(x,-k,t)\over 2}  , \quad\, \label{1}\\
&&\varepsilon\, {\partial F\over\partial x} = {e\over l}\, (n-N_{D}),  \label{2}\\   
&& n = { l\over 2\pi} \int_{-\pi/l}^{\pi/l} f(x,k,t) dk =
{ l\over 2\pi} \int_{-\pi/l}^{\pi/l} f^{FD}(k;n) dk,\quad \label{3}\\
&& f^{FD}(k;n) = {m^{*}k_{B}T\over\pi\hbar^{2}}\, \ln
\left[ 1 + \exp\left({\mu - E(k)\over k_{B}T}\right)\right] .
\label{4}
\end{eqnarray}  
Here $l$, $\varepsilon$, $f$, $n$, $N_{D}$, $k_{B}$, $T$, $-F$, $m^*$ and $-e<0$ are the 
SL period, the dielectric constant, the one-particle distribution function, the 2D electron 
density, the 2D doping density, the Boltzmann constant, the lattice temperature, the electric 
field, the effective mass of the electron, and the electron charge, respectively. The right side
of Eq.\ (\ref{1}) contains two collision terms. The first one has BGK 
(Bhatnagar-Gross-Krook) form, and it represents inelastic energy relaxation towards a 1D 
effective Fermi-Dirac distribution $f^{FD}(k;n)$ (local equilibrium) with collision 
frequency $\nu_{e}$. In Eq.\ (\ref{4}), the chemical potential $\mu$ depends on $n$ and 
is found by inverting the exact relation (\ref{3}). The second collision term accounts for 
impurity elastic collisions: 
\begin{eqnarray}
Q_{i}[f] &=& \nu_{i} {f(x,-k,t) - f(x,k,t)\over 2}\nonumber\\
&=& {l\over 2\pi} \int_{-\pi/l}^{\pi/l} [\overline{J}(k',k,x,t) - 
\overline{J}(k,k',x,t)]\, dk',  \label{5}\\
\overline{J}(k,k',x,t)&=& \nu_{i}\, {\pi\Delta |\sin kl|\over 2}
\delta(E(k) - E(k')) \, f(x,k,t),   \label{6}
\end{eqnarray}
provided we use the tight-binding miniband dispersion relation, $E(k)=\Delta\, 
(1-\cos kl)/2$ ($\Delta$ is the miniband width), and ignore transversal degrees of freedom. 
For simplicity, $\nu_{e}$ and $\nu_{i}$ will be fixed constants. The electron velocity in 
Eq.\ (\ref{1}) is $v(k) = E'(k)/\hbar = \Delta l\sin kl/(2\hbar)$.

To include shot-noise effects, we assume that the particle fluxes $\overline{J}(k,k',x,t)$
in the elastic collision term $Q_{i}[f]$ are replaced by fluctuating fluxes, $J(k,k',x,t)= 
\overline{J}(k,k',x,t)+ \delta J(k,k',x,t)$, where $\langle \delta J\rangle =0$. Two 
such fluxes are independent elementary processes unless their arguments are identical. For 
the same process, the correlations are those of a Poisson process, which yields 
\cite{BBupr00}:
\begin{eqnarray}
&&\langle \delta J(k_{1},k_{2},x,t)\delta J(k_{1}',k_{2}',x',t')\rangle = 
\nonumber\\
&& \quad {(2\pi)^2\over Al}\, \delta(k_{1}-k'_{1})\delta(k_{2}-k'_{2}) 
\delta(x-x')  \delta(t-t')\, \overline{J}(k_{1},k_{2},x,t),   \label{7}
\end{eqnarray}
in which $A$ is the area of the SL cross section. Inserting the fluctuating terms in Eq.\ 
(\ref{1}), we obtain a Boltzmann-Langevin equation (BLE) whose right side contains an 
additional term $\xi(x,k,t)$, with zero mean and correlation
\begin{eqnarray}
&& \langle \xi(x,t) \xi(x',t')\rangle = \delta(x-x') \delta(t-t')\, G(k,k',x,t), 
\label{8}\\
&& G(k,k',x,t) = {\pi\nu_{i}\over A}\, [f(x,k,t) + f(x,-k,t)]\, [\delta(k-k') - 
\delta(k+k')].       \label{9}
\end{eqnarray}

We shall now derive a reduced balance equation for the electric field. In order to do this, 
we shall assume that the electric field contribution in the BLE is comparable to the 
deterministic collision terms and that these terms dominate the other three. This is the 
so-called {\em hyperbolic limit}, in which the ratio of $\partial f/\partial t$ or $v(k)
\partial f/\partial x$ to $(eF/\hbar)\,\partial f/\partial k$ is of order $\epsilon
\ll 1$ \cite{BEPprb03}. Let $v_{M}$ and $F_{M}$ be electron velocity and field scales 
typical of the macroscopic phenomena described by the sought balance equation; for example, 
let them be the positive values at which the (zeroth order) drift velocity reaches its maximum. 
In the hyperbolic limit, the time $t_{0}$ it takes an electron with speed $v_{M}$ to traverse 
a distance $x_{0}=\varepsilon F_{M}l/(e N_{D})$, over which the field variation is of 
order $F_{M}$, is much longer than the mean free time between collisions, $\nu_{e}^{-1}
\sim \hbar/(eF_{M}l)=t_{1}$. We therefore define $\epsilon=t_{1}/t_{0}=\hbar v_{M}
N_{D}/(\varepsilon F_{M}^2 l^2)$ and formally multiply the fluctuating term $\xi$ and 
the two first terms on the left side of (\ref{1}) by $\epsilon$. After obtaining the number 
of desired terms, we set $\epsilon = 1$. The solution of Eq.\ (\ref{1}) for $\epsilon =0$ 
is straightforwardly calculated in terms of its Fourier coefficients as 
\begin{eqnarray}
&& f^{(0)}(k;n) = \sum_{j=-\infty}^{\infty} f^{(0)}_{j} e^{ijkl}\label{10}\\
&& f^{(0)}_{j} = {1-ij \varphi/\tau_{e}\over 1 + j^2 \varphi^{2}}
f^{FD}_{j}, \label{11}\\
&& \varphi = {F\over F_{M}}, \quad F_M = {\hbar\sqrt{\nu_{e}(\nu_{e}+\nu_{i})}
\over el}, \quad \tau_{e}= \sqrt{1+{\nu_{i}\over\nu_{e}}}. \label{12}
\end{eqnarray}
Note that Eq.\ (\ref{3}) implies $f^{(0)}_{0} = f^{FD}_{0} = n$. 
 
Once $f^{(0)}$ is known as a function of the unknown electron density and electric field, 
we make the Chapman-Enskog ansatz \cite{BEPprb03}:
\begin{eqnarray} 
&& f(x,k,t;\epsilon) = f^{(0)}(k;n) + \sum_{m=1}^{\infty} f^{(m)}(k;n)\, 
\epsilon^{m} ,    \label{13}\\
&&  {\partial n\over\partial t} = \sum_{m=0}^{\infty}  N^{(m)}(n)\, 
\epsilon^{m}.  \label{14}
\end{eqnarray} 
The coefficients $f^{(m)}(k;n)$ depend on the `slow variables' $x$ and $t$ only through 
their dependence on the electron density and the electric field (which is itself 
a functional of $n$). The electron density obeys a reduced evolution equation (\ref{14})
in which the functionals $N^{(m)}(n)$ are chosen so that the $f^{(m)}(k;n)$ are bounded 
and $2\pi/l$-periodic in $k$. Moreover the condition,
\begin{eqnarray} 
\int_{-\pi/l}^{\pi/l} f^{(m)}(k;n) \, dk = 2\pi\, f^{(m)}_{0}/l= 0,  \quad m\geq 1, 
\nonumber  
\end{eqnarray} 
ensures that $f^{(m)}$, $m\geq 1$, do not contain contributions proportional to the 
zero-order term $f^{(0)}$. $N^{(m)}(n)$ can be found by integrating (\ref{1}) over $k$,
using (\ref{3}), and inserting (\ref{13}) in the result: 
\begin{eqnarray} 
N^{(m)}(n) = - l\, {\partial\over\partial x}\int_{-\pi/l}^{\pi/l} v(k) 
f^{(m)}{dk\over 2\pi} .    \label{15}  
\end{eqnarray} 
Then integration of (\ref{2}) over $x$ yields a form of Amp\`ere's law:
\begin{eqnarray} 
\varepsilon {\partial F\over\partial t} + {e\over 2\pi}\,\sum_{m=0}^\infty
\epsilon^m\, \int_{-\pi/l}^{\pi/l} v(k)\, f^{(m)}(k;n) \, dk = J(t),
\label{16}  
\end{eqnarray} 
where $J(t)$ is the total current density. 

To find the equations for $f^{(m)}$, we insert Eqs.\ (\ref{13}) and (\ref{14}) in 
(\ref{1}), and then we equate like powers of $\epsilon$. The result is the following 
hierarchy of linear nonhomogeneous equations:
\begin{eqnarray} 
L f^{(1)} &=&  \left.  - \left({\partial \over \partial t} + v(k) {\partial \over 
\partial x}\right) f^{(0)}\right|_{0} + \xi^{(0)}(x,k,t), \label{17}\\
 L f^{(2)} &=&  \left.  \left. -  \left({\partial \over \partial t} 
 + v(k) {\partial \over \partial x}\right) f^{(1)}\right|_{0} - {\partial 
 \over \partial t} f^{(0)}\right|_{1}+ \xi^{(1)}(x,k,t), \quad \label{18}
\end{eqnarray} 
and so on. We have defined $L u(k) \equiv eF\hbar^{-1} du(k)/dk + ( \nu_{e} + 
\nu_{i}/2) u(k) + \nu_{i} u(-k)/2$, and the subscripts 0 and 1 mean that $\partial n/
\partial t$  is replaced by $N^{(0)}(n)$ and by $N^{(1)}(n)$, respectively. In Eq.\ 
(\ref{17}), the correlation of the noise $\xi^{(0)}(x,k,t)$ contains only the zeroth order 
distribution function. The linear equation $L u= S$  has a bounded $2\pi/l$-periodic solution 
provided $\int_{-\pi/l}^{\pi/l}S\, dk =0$. This solvability condition together with Eqs.\ 
 (\ref{17}), (\ref{18}), etc.\ also yield the previously found $N^{(m)}$ of Eq.\
 (\ref{15}) and the reduced equation (\ref{13}). 
 
\begin{figure}
\begin{center}
\includegraphics[width=12cm]{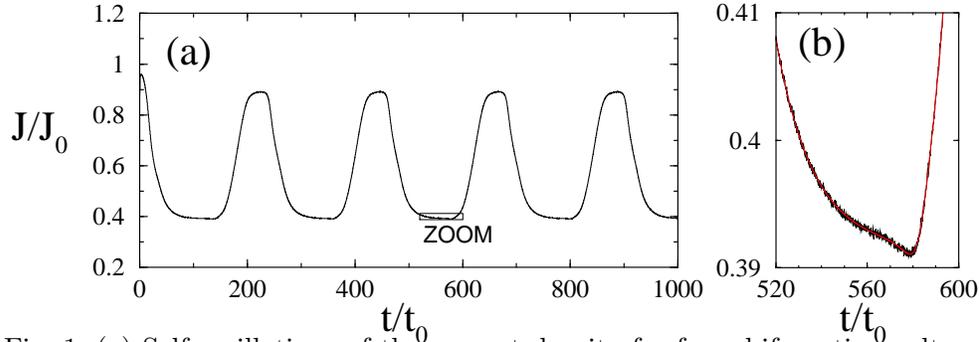}
\caption{(a) Self-oscillations of the current density far from bifurcation voltages. (b) 
Zoom of the time interval marked in (a) showing the effects of the shot noise. Data 
correspond to a 100-period GaAs/AlAs SL with well and barrier widths of 5.13 and 0.87 nm, 
respectively, and a 3D doping density of $1.4\times 10^{17}$ cm$^{-3}$, as in
\cite{SBHprb98}. The units of current density and time are $J_{0}= 1.9\times 10^5$ 
A/cm$^2$, $t_{0} = 9.7\times 10^{-14}$s, respectively, and the voltage is
115\% of the threshold voltage above which self-sustained oscillations appear. }
\label{fig1}
\end{center}
\end{figure}

\begin{figure}
\begin{center}
\includegraphics[width=12cm]{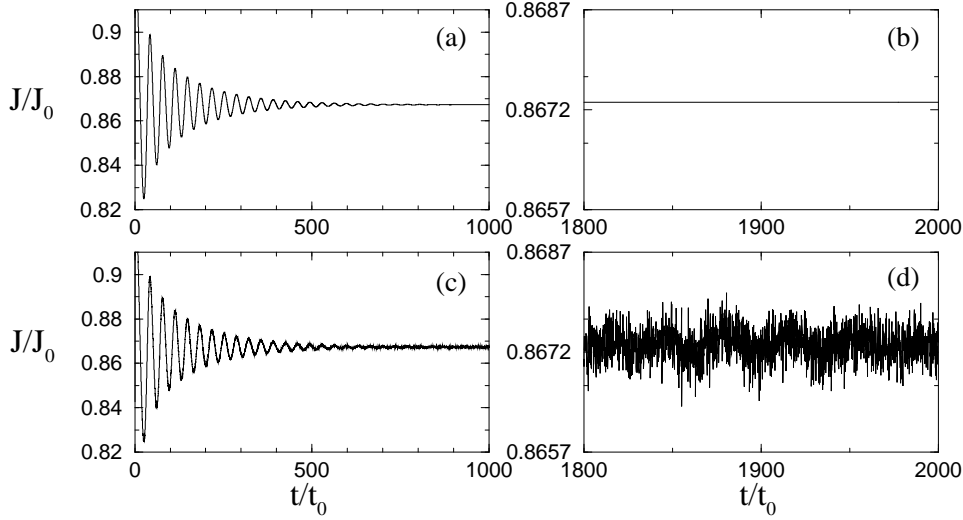}
\caption{Self-oscillations of the current density induced by the shot noise for 
voltages below the threshold voltage. (a) and (b) show the oscillatory relaxation
of the current towards its stationary value in the absence of noise. (c) and (d) show 
the self-sustained oscillations induced when the shot noise is present for the same
voltage value. Numerical values are as in Fig.~\ref{fig1}, but the voltage is
99.3\% of the threshold voltage above which self-sustained oscillations appear.}
\label{fig2}
\end{center}
\end{figure}

The solution of Eq.\ (\ref{17}) is 
\begin{eqnarray}
&& f^{(1)} = \nu_{e}^{-1}\, \sum_{j=-\infty}^{\infty} {\mbox{Re}S^{(1)}_{j} + 
i \tau_{e}^{-2}\mbox{Im}S^{(1)}_{j} - ij\varphi S^{(1)}_{j}/\tau_{e}\over 1 + 
j^2 \varphi^{2}}\, e^{ijkl},      \label{19}
\end{eqnarray} 
in which $S^{(1)}_{j}$ ($S^{(1)}_{0}=0$) is the $j$th Fourier coefficient of the right hand 
side of Eq.\ (\ref{17}). Using Eqs.\ (\ref{10}), (\ref{11}) and (\ref{19}), we 
explicitly write two terms in Eq.\ (\ref{16}), thereby obtaining (after setting $\epsilon=
1$) the following {\em stochastic generalized drift-diffusion equation} (SGDDE) for the 
electric field:
\begin{eqnarray}
&&\varepsilon {\partial F\over\partial t} + \tilde{v}\left(F,{\partial F\over
\partial x}\right)\, {eN_{D}\over l}\, \left( 1 + {\varepsilon l\over eN_{D}}\,
{\partial F\over\partial x}\right) - D\left(F,{\partial F\over\partial x}
\right)\,\varepsilon {\partial^2 F\over \partial x^2}\nonumber\\
&& \quad = - \delta J(x,t) + A\left(F,{\partial F\over\partial x}\right)\, J(t) ,   
\label{20}\\
&& A = 1 + {2 e v_{M} F_{M}^3\, [F_{M}^2 - (1+2 \tau^{2}_{e})\, F^2]\over 
\varepsilon l (\nu_{e}+ \nu_{i}) (F_{M}^2 + F^{2})^3 }\, n \tilde{m},  \label{21}\\
&& \tilde{v} = v_{M}V \tilde{m} \left( A - {\Delta B\over  2 e  }\, 
{\partial F\over\partial x} \right) ,   \label{22}\\
&& V(\varphi) = {2\varphi\over 1 + \varphi^2} , \quad
v_{M} = {\Delta l\, \tilde{I}_{1}(M)\over 4\hbar\tau_{e} \tilde{I}_{0}(M)},  
\label{23}\\
&& \tilde{I}_{m}(s) = \int_{-\pi}^{\pi}\cos (m k)\,\ln\left( 1+e^{s-\delta+
\delta\cos k}\right)\,dk, \label{24}\\
&& \tilde{m}\left({n\over N_{D}}\right) = {\tilde{I}_{1}(\mu/k_{B}T)\, \tilde
{I}_{0}(M)\over \tilde{I}_{1}(M) \tilde{I}_{0}(\mu/k_{B}T)}, \label{25}\\
&& \tilde{m}_{2}\left({n\over N_{D}}\right) = {\tilde{I}_{2}(\mu/k_{B}T)\, 
\tilde{I}_{0}(M)\over \tilde{I}_{1}(M) \tilde{I}_{0}(\mu/k_{B}T)}, \label{26}\\
&& D = {\Delta^2 l F_{M}\over 8\hbar e\tau_{e} \, (F^2_{M} + F^{2}) }  
 \left( 1 - {4\hbar v_{M}C\over\Delta l} \right) ,  \label{27}\\
&& B = { (5F^2_{M}-4F^2) \tilde{m}_{2}\over (F^2_{M}+ 4F^2)^2 \tilde{m}} 
 - {4\hbar v_{M} F^2_{M}(F^2_{M}-F^2) (\tau_{e} + \tau^{-1}_{e}) 
(n\tilde{m})' \over \Delta l (F^2_{M}+F^2)^3}, \label{28}\\
&& C = {\tau_{e}  (F^2_{M}-2F^2)  (n \tilde{m}_{2})' \over F^2_{M} +  4 F^2 }
+ {8\hbar v_{M} (1+\tau_{e}^{2})[F F_{M}(n \tilde{m})']^2\over \Delta l 
( F^2_{M}+F^2)^2}.  \quad \quad   \label{29}
\end{eqnarray} 
Here the electron density is given by the Poisson equation (\ref{2}), $\delta=\Delta/(
2k_{B}T)$ and $g'$ means $dg/dn$. The fluctuating current density $\delta J(x,t)$ in 
Eq.\ (\ref{20}) has zero mean and correlation
\begin{eqnarray}
\langle\delta J(x,t)\delta J(x',t')\rangle = {e^2\Delta^2\nu_{i}l\,
\delta(x-x') \delta(t-t') \over 8\pi A\hbar^2\nu_{e}(\nu_{e}+\nu_{i}) 
(1+\varphi^2)^2} \left(n- {4\hbar\tau_{e}v_{M} n\tilde{m}_{2}\over 
\Delta l (1+4\varphi^2)}\right). \quad \,    \label{30}
\end{eqnarray} 
The effects of shot noise on the solutions of the deterministic equation are more
noticeable near bifurcation points. Thus they should affect more evidently the 
onset of the self-sustained oscillations of the current that occur in SCSLs 
\cite{SBHprb98}. These stable time-periodic solutions of the deterministic model 
are due to periodic charge dipole motion from one contact to the other and their 
frequency agrees with experimental observations \cite{BEPprb03}. Far from the 
bifurcation voltages, the effect of shot noise is small, as shown in Fig.~\ref{fig1}. 
However, near the voltage value at which self-oscillations start in the deterministic
model, the effects of shot noise are more noticeable. Fig.~\ref{fig2} shows that the 
shot noise may induce self-oscillations of the current at voltage values below threshold,
for which the deterministic model shows relaxation towards the stationary current.
A great enhancement in the variance of fluctuations near threshold should also 
occur, similarly to the effects of noise on the onset of Gunn oscillations in bulk 
semiconductors \cite{Gunssc63}, cf.\ Section 7.6 of \cite{Keibook87}. 

\section{Shot-noise effects in weakly-coupled superlattices}
\label{sec:WCSL}
Deterministic charge transport in weakly coupled SLs is described by discrete balance 
equations \cite{Bonjpcm02}. To this day, no one has derived these equations from 
quantum kinetic theory, and therefore we cannot study the effects of shot noise on
discrete balance equations by perturbative methods similar to those outlined in Section
\ref{sec:SCSL}. What we can do is to add fluctuating particle fluxes to the discrete
balance equations and assume Poissonian statistics for them. 

Let us consider a WCSL with $N+1$ barriers and $N$ wells. The zeroth barrier separates the
injecting region from the first SL well, whereas the $N$th barrier separates the $N$th well 
from the collecting region. Assume that $F_{i}$ is the average electric field across one SL 
period, consisting of the $i$th well and the $(i-1)$th barrier. Similarly, $n_{i}$ is the 2D 
electron density at the $i$th well, concentrated in a plane perpendicular to the growth 
direction inside the $i$th well. Then, the Poisson equation (averaged over the $i$th period) 
and the charge continuity equation are
\begin{eqnarray}
&& F_{i}-F_{i-1} = {e\over\varepsilon}\, (n_{i}-N_{D}) .
\label{eq2c1}\\
&& e\, {dn_{i}\over dt} = J_{i-1\to i} - J_{i\to i+1}. \label{eq2c2}
\end{eqnarray}
Here $J_{i\to i+1}$ is the tunneling current density across the $i$th barrier, i.~e., from well 
$i$ to well $i+1$ with $i=1,\ldots, N$. Assuming that the intersubband scattering times are
much smaller than the escape time from a well, which, in turn, is much smaller than the
dielectric relaxation time, the main contributions to $J_{i\to i+1}$ are due to sequential 
resonant tunneling. Electrons in the lowest subband $E_{1}$ of the $i$th well tunnel to one
of the excited levels of the $(i+1)$th well, and then immediately relax to the lowest level
thereof. In the limit of small $\gamma_{\nu}$ (broadening of subbands due to scattering), 
the following expression for the {\em deteministic} tunneling current, 
$\overline{J}_{i\to i+1}$, holds \cite{Bonjpcm02}:
\begin{eqnarray}
&& \overline{J}_{i\to i+1} = {e\, v^{(f)}(F_i)\over l}\,  \left\{ n_i - {m^{*}
k_{B}T\over \pi\hbar^{2}}\, \ln
\left[1+ e^{-{eF_i l\over  k_{B}T}}\left( e^{\frac{\pi\hbar^{2}
n_{i+1}}{m^{*}k_{B}T}} -1\right)\right] \right\} ,\label{eq2c3}\\
&& v^{(f)}(F_i ) = \sum_{\nu=1}^{n_{max}} \frac{\hbar^{3} l}{2 m^{* 2}}\,
{(\gamma_{1}+\gamma_{\nu}) T_i(E_{1})\over (E_{1}
-E_{\nu}+ eF_i l)^{2} + (\gamma_{1}+\gamma_{\nu})^{2}}. \label{eq2c4}
\end{eqnarray}
The {\em forward drift velocity} $v^{(f)}(F_i )$ is a sum of Lorentzians centered at the 
resonant field values $F_i = (E_{\nu} - E_{1})/(el)$. $T_{i}(E_{1})$ is proportional to the 
transmission coefficient through the $i$th barrier. The tunneling current is a linear function 
of $n_{i}$, but it is a strongly nonlinear function of $n_{i+1}$. Moreover, $J_{i\to i+1} 
\sim e v^{(f)}(F_i ) n_{i}/l$, for $F_{i}$ of the order of the first resonant value or larger. 
For such values, the resulting tunneling current density has the same shape as assumed in the 
original {\em discrete drift model} \cite{BGCprb94}. 

Differentiating Eq.~(\ref{eq2c1}) with respect to time and inserting the result into 
Eq.~(\ref{eq2c2}), we obtain
\begin{eqnarray}
\varepsilon {dF_{i}\over dt} + J_{i\to i+1} = J(t),   \label{eq2c5}
\end{eqnarray}
in which the total current density $J(t)$ is the same function for all $i$. 
Equation~(\ref{eq2c5}) is a discrete version of Amp\`ere's equation. We complete our 
description by adding the voltage bias condition
\begin{eqnarray}
{1\over (N+1)}\, \sum_{i=0}^{N} F_i &=& {V(t)\over (N+1)l}\,, \label{eq2c6}
\end{eqnarray}
for the known voltage $V(t)$. 

Our discrete system of deterministic equations consists of Eq.~(\ref{eq2c1}) for $i=1,
\ldots,N$, Eq.~(\ref{eq2c5}) for $i=0,\ldots,N$, and Eq.~(\ref{eq2c6}). In total, we 
have $(2N+2)$ equations for the unknowns $n_{1}$, \ldots, $n_{N}$, $F_{0}$, \ldots, 
$F_{N}$, and $J(t)$. We need to specify the constitutive relations $J_{0\to 1}$ (tunneling 
from the injecting region to the SL), and $J_{N\to N+1}$ (tunneling from the SL to the 
collecting region). Equation~(\ref{eq2c5}) evaluated for $i=0$ and $i=N$ determines the 
boundary conditions. As tunneling currents at the boundaries, we use the following 
phenomenological expressions \cite{AWBpre01}:
\begin{eqnarray}
\overline{J}_{0\to 1} = \sigma F_{0}, \quad
\overline{J}_{N \to N+1}= {n_{N} \sigma F_{N}\over N_{D}}\,.    \label{eq2c7}
\end{eqnarray}
These conditions are particular cases of the tunneling currents described in 
\cite{Bonjpcm02}. The deterministic model consists of Eqs.~(\ref{eq2c1}) and 
(\ref{eq2c3}) -- (\ref{eq2c7}). 

In the stochastic model, zero-mean random currents $\delta J_{i\to i+1}(t)$ have to be 
added to the tunneling current densities $\overline{J}_{i\to i+1}$: 
\begin{eqnarray}
J_{i\to i+1} = \overline{J}_{i\to i+1} + \delta J_{i\to i+1}(t).   \label{eq2c8} 
\end{eqnarray}
The currents $\delta J_{i\to i+1}$ have correlations \cite{BSSprb02}
\begin{eqnarray}
&&\langle \delta J_{i \to i+1}(t)\, \delta J_{j \to j+1}(t')\rangle =
{e^2 v^{(f)}(F_i)\delta_{ij} \delta(t-t')\over Al} \nonumber\\
&& \quad \quad \times  \left\{ n_i  + {m^{*}
k_{B}T\over \pi\hbar^{2}}\, \ln
\left[1+ e^{-{eF_i l\over  k_{B}T}}\left( e^{\frac{\pi\hbar^{2}
n_{i+1}}{m^{*}k_{B}T}} -1\right)\right]\right\},\quad  \, \label{eq2c9}\\
&& \langle \delta J_{0 \to 1}(t)\, \delta J_{0 \to 1}(t')\rangle =
{e\sigma\, F_0 \over A}\, \delta(t-t')  , \label{eq2c10} \\
&& \langle\delta J_{N \to N+1}(t) \delta J_{N\to N+1}(t')\rangle = {e n_{N}
\sigma F_N\over A N_{D}}\, \delta(t-t'),  \label{eq2c11}
\end{eqnarray}
for $i=1,\ldots, N-1$. The idea behind this form of random tunneling current is that 
uncorrelated electrons are arriving at the $i$-th barrier with a distribution function of time 
intervals between arrival times that is Poissonian \cite{BBupr00}. Moreover, the correlation 
time is of the same order as the tunneling time, so that it is negligible on the longer time scale 
of dielectric relaxation. 

The high-temperature limit of the stochastic model has been numerically solved in 
\cite{BSSprb02}. Here we shall study a particular effect of shot noise on the nonlinear 
dynamics of a WCSL. The current-voltage characteristics of the corresponding deterministic 
model consist of a number of stable solution branches corresponding to electric 
field profiles displaying two domains separated by a domain wall. The current-voltage
plane corresponds to the usual bifurcation diagram of a norm of a solution (the total 
current density) versus the bifurcation parameter (the voltage). We should note that two
neighboring stable solution branches are typically connected via an intermediate unstable 
solution branch at two limit points that are saddle-node bifurcations. If the initial state 
corresponds to a voltage on the middle of a stable branch, $V_{i}$, and is suddenly 
switched to a final voltage corresponding to the next branch, $V_{f}$, the domains 
relocate after a certain delay time, called relocation time \cite{AWBpre01}. 

The possible scalings of this mean relocation time can be calculated by means of 
bifurcation theory and the classical results for escape of a Brownian particle from a 
potential well. The idea is as follows. Let $(V_{L}, J_{L})$ be the limit point
connecting one stable and one unstable branch of deterministic stationary solutions via 
a saddle-node bifurcation: two solutions exist for $V<V_{L}$, and they disappear for
$V>V_{L}$. Let us assume that $V_{i}<V_{f}<V_{L}$, and that for $V=V_{f}$ three
stationary solutions coexist with current densities $J_{1}<J_{2}<J_{3}$. The solution 
branch corresponding $J=J_{3}$ is the same stable branch containing the initial point. This
branch coalesces with the branch of unstable solutions corresponding to $J_{2}$ at the limit
point $(V_{L}, J_{L})$. Provided $(V_{f}-V_{L})$ is of the same order as the noise 
amplitude in the appropriate dimensionless units, a projection of the noise on the eigenvector 
corresponding to the zero eigenvalue in the linear stability problem at $V_{f}=V_{L}$ 
appears in the amplitude equation for the saddle-node bifurcation. The latter becomes the 
following stochastic amplitude equation:
\begin{equation}
{da\over dt} = \alpha\, (V_{f}-V_{L}) + \beta a^2 + \sqrt{2\gamma}\, \xi(t). 
\label{eq2c12}
\end{equation}
In this expression, $a$ is proportional to $|J-J_{L}|$ and $\xi(t)$ is the zero-mean, 
delta-correlated white noise. The positive parameters $\alpha$, $\beta$, and $\gamma$ 
can be calculated by projection methods \cite{IJobook80}, but their precise form does not 
matter for the argument we want to make. Note that Eq.~(\ref{eq2c12}) has two stationary 
solutions with $|J-J_{L}|\propto |V_{f}-V_{L}|^{1/2}$ if $\gamma=0$. 

For $V_{f}<V_{L}$, Eq.~(\ref{eq2c12}) describes the escape of a Brownian particle 
from a potential well corresponding to the cubic potential $U(a;V_{f}-V_{L})= \alpha\, 
(V_{f}-V_{L})\, a + \beta a^3/3$. Provided the height of the barrier is large compared 
to the noise strength, the reciprocal of the mean escape time is proportional to the
equilibrium probability density $P= e^{-U/\gamma}/Z$ evaluated at the maximum of the 
potential. This yields
\begin{equation}
\tau_{reloc} \sim {\pi\over\sqrt{\alpha\beta\,|V_{f}-V_{L}|}} 
\exp\left({4\, |\alpha(V_{f}-V_{L})|^{3/2}\over 3\gamma\sqrt{\beta}}
\right),    \label{eq2c13}
\end{equation}
cf.\ Eq.\ (8.6.17) in \cite{Schbook80}. Thus, there exists a relatively large voltage 
interval $|V_{f}-V_{L}|^{3/2}\gg \gamma\beta^{1/2}\alpha^{-3/2}$, over which the 
logarithm of the relocation time scales superlinearly with $|V_{f}-V_{L}|$. In terms
of the current, ln$(\tau_{reloc}) \propto |J-J_{L}|^3$, because $a\propto |J-J_{L}|
= O(|V_{f}-V_{L}|^{1/2})$. 

There is some evidence that the parabolic region near the limit point might be  
so narrow that the amplitude equation (\ref{eq2c12}) breaks down. For example, 
Fig.~5 of Ref.~\cite{APMprb97} shows that the maximum and the limit point of the 
current-voltage characteristic are extremely close for all branches. In this case, it may 
very well happen that the region $\gamma^{2/3}\beta^{1/3}/\alpha\ll |V_{f}-V_{L}|
\ll K$ in which Eq.\ (\ref{eq2c12}) holds is too narrow. This case might be
better approximated by analyzing the effect of shot noise on a spiky limit point, at 
which the slope of the current-voltage characteristic is discontinuous, and the term
proportional to $a^2$ in (\ref{eq2c12}) is replaced by a piecewise linear function of $a$. 
Then a simple calculation of the barrier height yields $\ln(\tau_{reloc})\propto |V_{f}-
V_{L}|^{2}\propto |J-J_{L}|^{2}$ instead of Eq.\ (\ref{eq2c13}). The numerical 
solution of the stochastic equations could be used to discriminate between this prediction
and Eq.\ (\ref{eq2c13}). There are early experiments in which  ln$(\tau_{reloc})$
was fitted to a straight line \cite{LGPprb98}, but no theoretical prediction existed at that 
time and perhaps more careful measurements should be attempted in order to validate
the theory. 

\section{Conclusions}
\label{sec:fin}
We have modelled the effects of shot noise in strongly and weakly coupled superlattices.
For a SCSL, we have proposed a Boltzmann-Langevin equation containing fluctuating
terms that represent shot noise. Then we have used a consistent Chapman-Enskog method to 
derive a stochastic drift-diffusion equation for the electric field. In the case of WCSLs,
we have slightly generalized the discrete Langevin equations proposed in \cite{BSSprb02},
and found the scaling for the mean relocation time that electric field domains take to
react to a switch in the applied voltage.

\bigskip
\bigskip

The figures in this paper were calculated by Guido Dell'Acqua and Ram\'on Escobedo, 
to whom I am very much indebted. This work has been supported by the MCyT grant 
BFM2002-04127-C02, and by the European Union under grant HPRN-CT-2002-00282.

\end{document}